\documentclass[preprint]{aastex}

\shorttitle{$Fermi$ LAT Observation of a Gamma-ray Source at Eta Car}
\shortauthors{$Fermi$ collaboration et al.}

\begin{document}


\title{$Fermi$ Large Area Telescope Observation of a Gamma-ray Source at the Position of Eta Carinae}



\author{
A.~A.~Abdo\altaffilmark{1,2}, 
M.~Ackermann\altaffilmark{3}, 
M.~Ajello\altaffilmark{3}, 
A.~Allafort\altaffilmark{3}, 
L.~Baldini\altaffilmark{4}, 
J.~Ballet\altaffilmark{5}, 
G.~Barbiellini\altaffilmark{6,7}, 
D.~Bastieri\altaffilmark{8,9}, 
K.~Bechtol\altaffilmark{3}, 
R.~Bellazzini\altaffilmark{4}
% ...
}
\altaffiltext{1}{Space Science Division, Naval Research Laboratory, Washington, DC 20375, USA}
\altaffiltext{2}{National Research Council Research Associate, National Academy of Sciences, Washington, DC 20001, USA}
\altaffiltext{3}{W. W. Hansen Experimental Physics Laboratory, Kavli Institute for Particle Astrophysics and Cosmology, Department of Physics and SLAC National Accelerator Laboratory, Stanford University, Stanford, CA 94305, USA}
\altaffiltext{4}{Istituto Nazionale di Fisica Nucleare, Sezione di Pisa, I-56127 Pisa, Italy}
\altaffiltext{5}{Laboratoire AIM, CEA-IRFU/CNRS/Universit\'e Paris Diderot, Service d'Astrophysique, CEA Saclay, 91191 Gif sur Yvette, France}
\altaffiltext{6}{Istituto Nazionale di Fisica Nucleare, Sezione di Trieste, I-34127 Trieste, Italy}
\altaffiltext{7}{Dipartimento di Fisica, Universit\`a di Trieste, I-34127 Trieste, Italy}
\altaffiltext{8}{Istituto Nazionale di Fisica Nucleare, Sezione di Padova, I-35131 Padova, Italy}
\altaffiltext{9}{Dipartimento di Fisica ``G. Galilei", Universit\`a di Padova, I-35131 Padova, Italy}
% ...

%
%
%




\begin{abstract}

The Large Area Telescope (LAT) onboard the $Fermi$ Gamma-ray Space Telescope detected
a $\gamma$-ray source that is
spatially consistent with the location of Eta Carinae
This source has been persistently bright since the beginning of the LAT survey observations (from 2008 August to 2009 July, the time interval considered here).
The $\gamma$-ray signal is detected significantly throughout the LAT energy band (i.e., up to $\sim$100 GeV).
The 0.1--100 GeV energy spectrum is well represented
by a combination of a cutoff power-law model ($<$ 10 GeV) and a hard power-law component ($>$ 10 GeV).
The total flux ($>$ 100 MeV) is $3.7^{+0.3}_{-0.1} \times 10^{-7}$ photons s$^{-1}$ cm$^{-2}$, with additional systematic uncertainties of 10\%,
and consistent with the average flux measured by $AGILE$ \citep{agile}.
The light curve obtained by $Fermi$ is consistent with steady emission.
Our observations do not confirm the presence of a $\gamma$-ray flare in
2008 October as reported by \citet{agile},
although we cannot exclude that a flare lasting only a few hours escaped
detection by the $Fermi$ LAT.
We also do not find any evidence for $\gamma$-ray variability
that correlates with the large X-ray variability of Eta Carinae
observed during 2008 December and 2009 January.
We are thus not able to establish an unambiguous identification of the LAT source with Eta Carinae.

\end{abstract}




\section{Introduction}

Eta Carinae (hereafter $\eta$ Car), with its famous bipolar nebula, 
is in the Carina nebula at a distance of 2.3 kpc \citep{davidson}.
The star is a luminous blue variable which is thought to be near the onset of pulsational instabilities.
Since the flux from radio to X-ray is known to vary with a period of 5.54 yr \citep{cycle},
$\eta$ Car is considered to have a companion star and hence be in a binary system.
Around periastron, the 2--10 keV X-ray light curve first increases gradually and then drops sharply to the minimum by more than one order of magnitude \citep{hamaguchi}.
After the faintest state lasting a few months,
the source returns to the nominal flux.
The last periastron passage was in 2008 December -- 2009 January and the X-ray flux behaved in a similar fashion as observed in previous orbits \citep{rxte_lc}.

The binary system parameters of $\eta$ Car are currently estimated as follows: stellar masses 
$M_{\eta} \sim 120 M_{\odot}$ and $M_{\rm c} \sim 30  M_{\odot}$
with an eccentricity of 0.9--0.95,
mass-loss rates and terminal velocities of the stellar winds
$\dot{M}_{\eta} \sim 10^{-4} M_{\odot}$ yr$^{-1}$, $\dot{M}_{\rm c} \sim 10^{-5} M_{\odot}$ yr$^{-1}$,
$v_{\eta,\infty} \sim 500$ km s$^{-1}$ and $v_{{\rm c},\infty} \sim 3000$ km s$^{-1}$ ,
where the subscripts $_{\eta}$ and $_{\rm c}$ represent $\eta$ Car and the companion, respectively \citep[where references are given]{parameter}.
~Such binary systems with stellar winds at high mass-loss rates and velocities are called colliding wind binaries (CWBs),
and are theoretically predicted to be potential sites of high-energy $\gamma$-ray
emission through strong shocks due to the colliding winds \citep{cwb},
although there have been no distinct observational detections of $\gamma$-rays  from CWBs so far.

Recently, two $\gamma$-ray satellites, $AGILE$ and $Fermi$ discovered a $\gamma$-ray source that is spatially consistent with $\eta$ Car \citep{agile,lat_bright}.
The source also overlaps with the error circles of the unidentified
COS-B and EGRET sources, 2CG 288--00 \citep{cosb} and 3EG J1048--584 \citep{3eg}.
If this source were indeed associated with $\eta$ Car,
it would be the first time that a CWB is observed in high-energy $\gamma$-rays.
\citet{agile} reported the $\gamma$-ray average flux ($>$ 100 MeV) of $ 3.7 \pm 0.5 \times 10^{-7}$ photons s$^{-1}$ cm$^{-2}$ and an increase of $\gamma$-ray flux 5--9 times above the average on a time scale of 2 days in 2008 October.
In this letter, we analyze the $\gamma$-ray source
observed by the Large Area Telescope (LAT) onboard $Fermi$
covering the recent periastron passage,
and discuss the possibility of the association with $\eta$ Car.
Hereafter, we refer to the $\gamma$-ray source as 1FGL J1045.2--5942 as quoted in the $Fermi$ LAT first source catalog \citep{lat_1yr}.

\section{Data Reduction and Results}

The $Fermi$ Gamma-ray Space Telescope was launched on 2008 June 11th
and the LAT instrument began nominal sky-survey observations on 2008 August 4th.
In this survey mode, the LAT scans the sky completely every 2 orbits ($\sim$3 hours);
details of the LAT performance are described in \citet{lat_performance}.
To study the $\gamma$-ray spectrum and light curve of 1FGL J1045.2--5942,
here we analyze the LAT data observed from 2008 August 4th to 2009 July 23rd,
with the standard analysis software of Science Tools version v9r11
and the instrument response function of P6\_V3\_DIFFUSE.
To minimize background,
we used only the ``diffuse'' class events
(events which have the highest probability of being photons),
and require the
reconstructed zenith angles to be less than $105\degr$ in order to exclude Earth's albedo photons.
Within the radius of 10$\degr$ around 1FGL J1045.2--5942,
there are no $\gamma$-ray bursts detected by the LAT during this exposure.
The systematic uncertainty of the effective area of
the response function (P6\_V3\_DIFFUSE) is estimated
to be 10\%, 5\% and 20\% at 100 MeV, 560 MeV and 10 GeV, respectively.
In the following analysis,
we also consider the $\gamma$-ray emission from the pulsars
PSR J1048--5832 and PSR J1028--5819
[see Abdo et al.~(2009b; 2009e) for details of these pulsars]
as a comparison.
These pulsars are only $\sim1\degr$ and $\sim2.5\degr$ away from 1FGL J1045.2--5942 and like most pulsars, display no significant variability (except for their pulsations).
The comparison of the light curves and spectra for the three sources provides an estimate of their relative variabilities or differences, regardless of systematic uncertainties.

Figure~\ref{fig1} shows counts maps around 1FGL J1045.2--5942
for the energy bands of 0.1--10 GeV and 10--100 GeV.
In addition to the bright pulsars PSR J1048--5832 and PSR J1028--5819 in the soft energy band,
1FGL J1045.2--5942 is clearly detected in both soft and hard energy bands
and is the brightest $\gamma$-ray source above 10 GeV in this region.
Applying the source-finding tool ($\tt gtfindsrc$) for the total 0.1--100 GeV band,
the position (J2000) of 1FGL J1045.2--5942 is obtained as 
$(\alpha, \delta) = (161.281, -59.710)$
with the 95\% statistic error of $0{\degr}.025$.
This result is consistent with the $Fermi$ LAT first source catalog \citep{lat_1yr},
and
places $\eta$ Car slightly outside the 95\% error region (offset $0{\degr}.03$).
However, when we perform the same analyses
for the soft and hard bands separately,
the source position is estimated as
$(\alpha, \delta) = (161.312, -59.713)$ with the 95\% error of $0{\degr}.062$ and $(161.265, -59.695)$ with the error of $0{\degr}.031$, respectively.
The hard band position is offset by only $0{\degr}.01$ from the location of $\eta$ Car and
thus is fully consistent with the object. The soft band position is offset by $0{\degr}.04$
from $\eta$ Car, yet due to the larger uncertainty in the soft band we cannot
conclude whether the LAT indeed detects two distinct sources or whether the soft and
hard components arise from the same object.
On the other hand, despite the many interesting nearby sources [the hard X-ray source IGR J10447--6027 and the anomalous X-ray pulsar 1E 1048.1--5937 that have been detected by ${\it INTEGRAL}$ \citep{integral} and other bright X-ray sources (WR 25, HD 93129A/B, HD 93250, HD 93205 and HD 93160) as well as the recently-discovered neutron star \citep{x-ray_pulsar}],
none of these fall in the 95\% LAT error circle.

\subsection{Spectral Analysis}

To measure the time-averaged energy spectrum of 1FGL J1045.2--5942,
we performed a spectral analysis
with the maximum likelihood fitting tool ($\tt gtlike$).
We selected a region of interest (ROI) with radius of $10\degr$ around the source and modeled the observed $\gamma$-ray emission with a combination of point sources, Galactic and extragalactic diffuse emission, and residual instrumental background.
Within the ROI, we found 25 point-like sources of which 4 have been identified as pulsars \citep{lat_1yr}.
We modeled the spectra of the pulsars using
an exponentially cut-off power-law (CPL) model,
$E^{-{\it \Gamma}}$exp($-E/E_{\rm cutoff})$,
while the other sources are modeled using a simple power-law (PL) model, $E^{-{\it \Gamma}}$,
where $E$ is the photon energy, ${\it \Gamma}$ is the photon index
and $E_{\rm cutoff}$ is the cutoff energy.
The Galactic diffuse emission was modeled using GALPROP
\citep[e.g.,][]{galprop},
version ``gll\_iem\_v01.fit'',
and the extragalactic diffuse
emission and residual instrumental backgrounds were combined into a spatially uniform component with a power-law spectral model.
\footnote{A check with the more-recent LAT-team model for diffuse emission, gll\_iem\_v02.fit, and isotropic spectrum isotropic\_iem\_v02.txt yielded entirely consistent results.}

\begin{figure}
\includegraphics[width=0.7\textwidth]{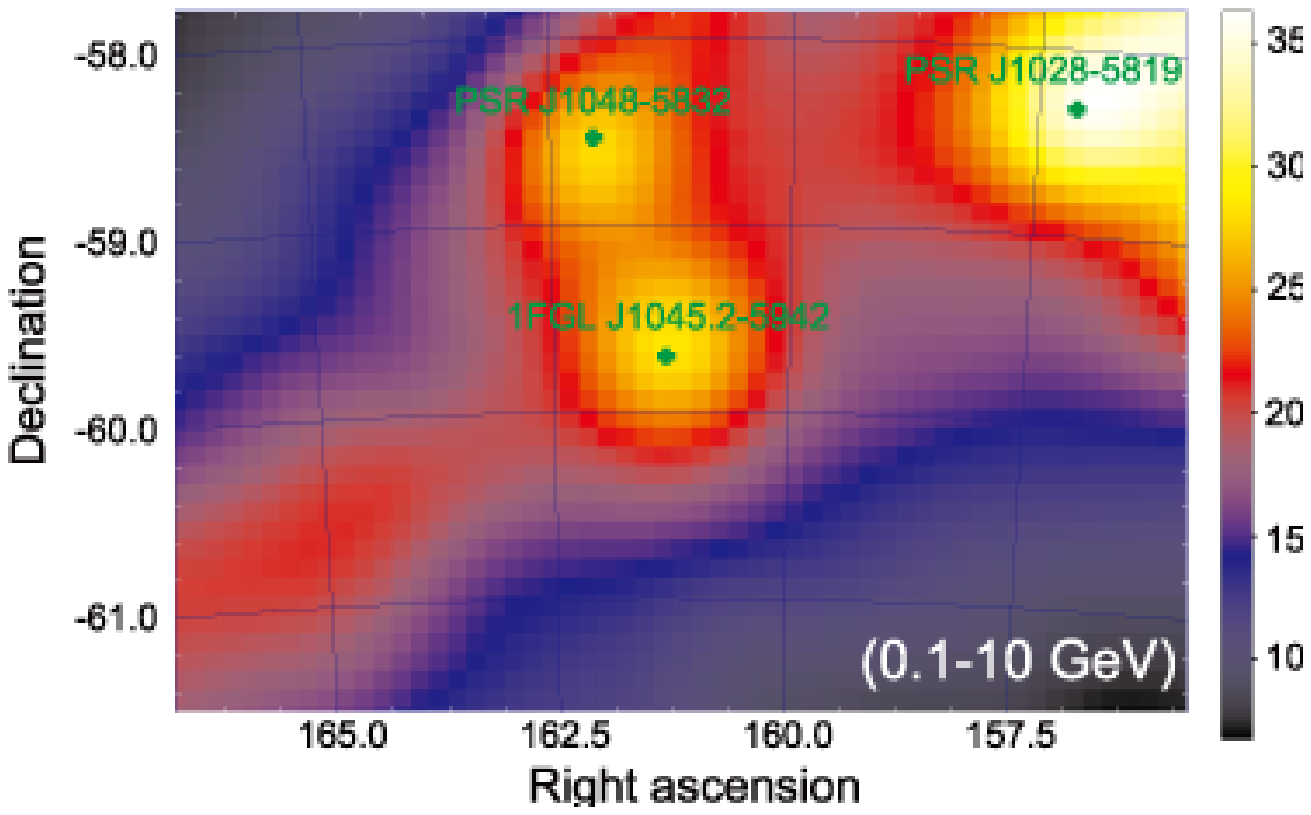}
\includegraphics[width=0.7\textwidth]{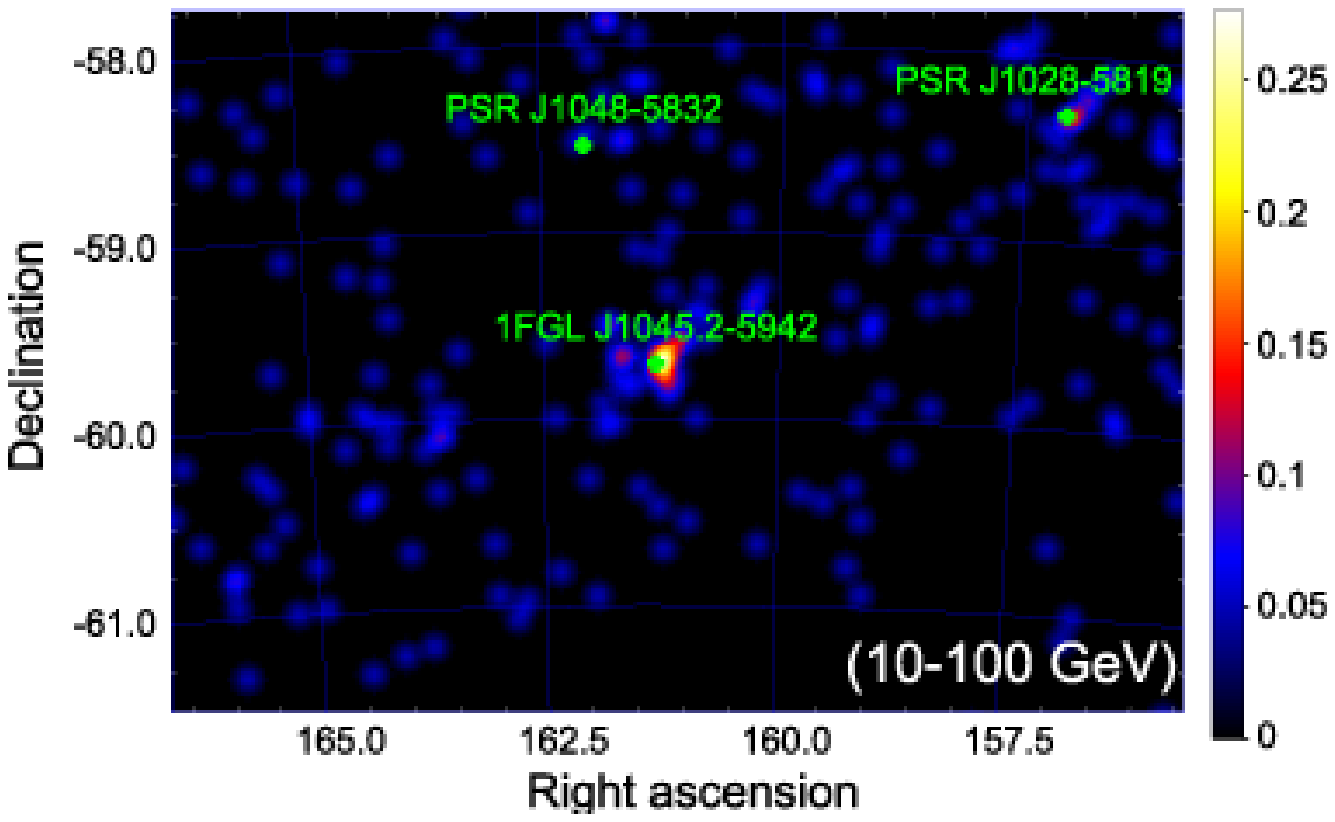}
\caption{Counts maps around 1FGL J1045.2--5942
with the energy bands of 0.1--10 GeV (top)
and 10--100 GeV (bottom),
where the pixel sizes are $0{\degr}.1$ with 6-bin Gaussian smoothing and $0{\degr}.025$ with 4-bin smoothing, respectively.
The unit of the color bars is counts pixel$^{-2}$.
Green crosses indicate the positions of the $Fermi$ LAT first source catalog \citep{lat_1yr}.}
\label{fig1}
\end{figure}

\begin{figure}
\includegraphics[width=0.5\textwidth]{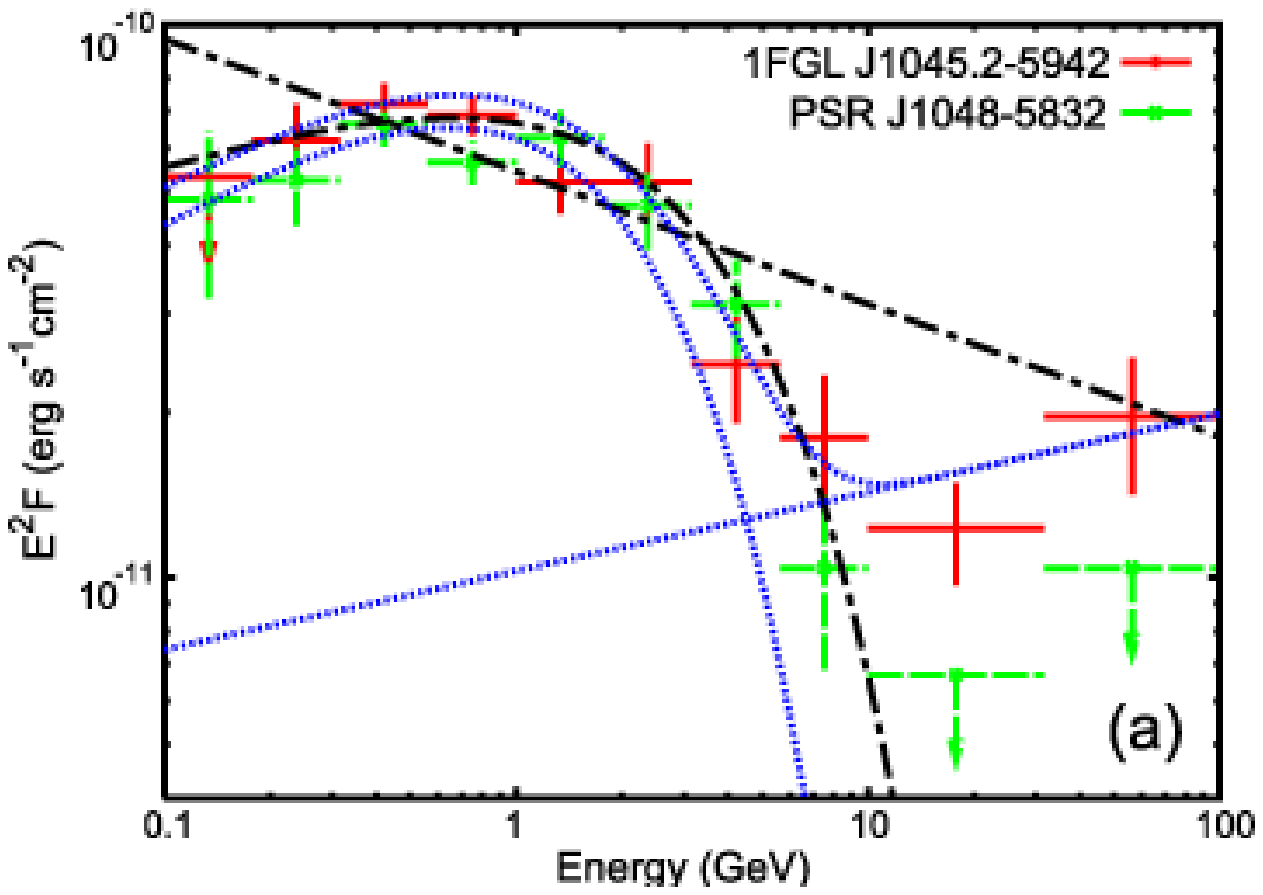}
\includegraphics[width=0.5\textwidth]{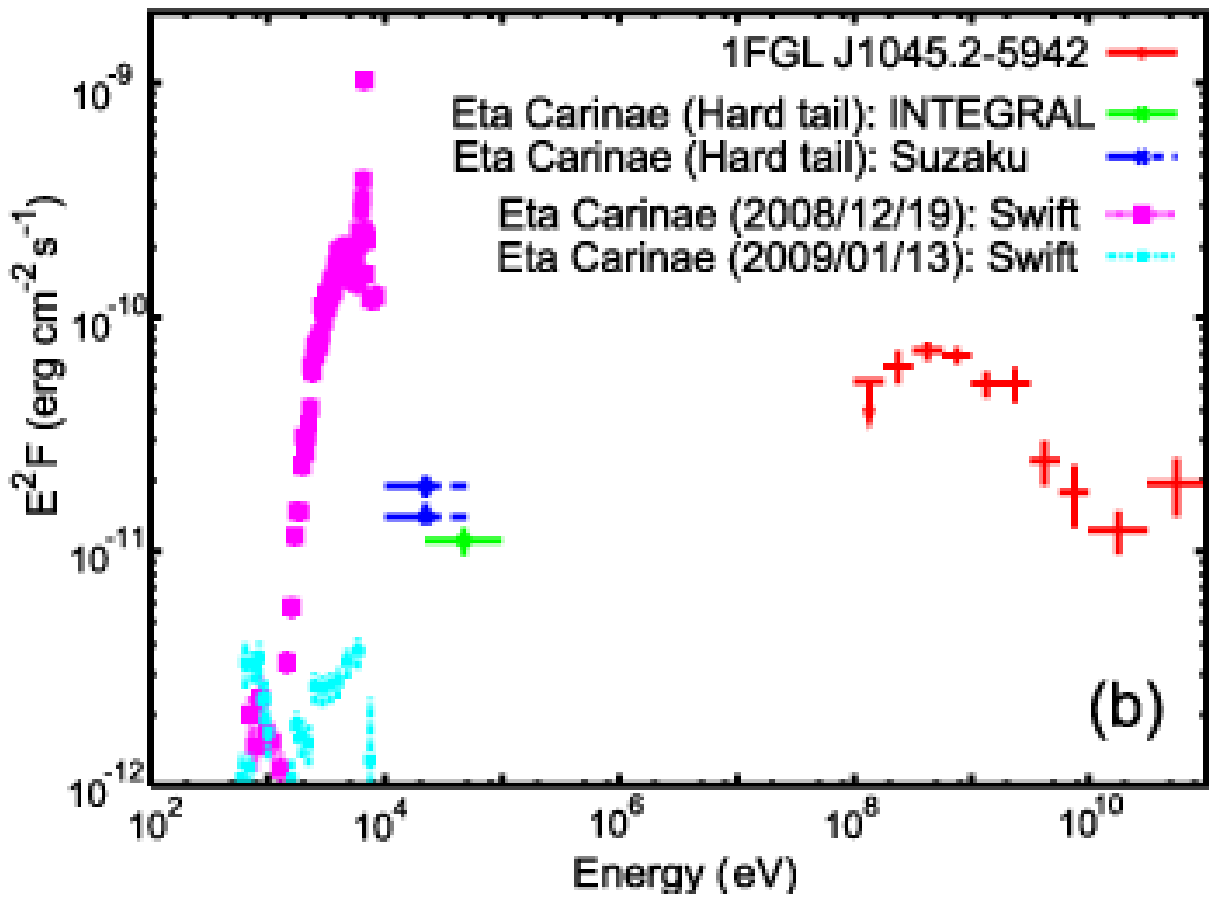}
\caption{(a) Spectral energy distributions of 1FGL J1045.2--5942 (red solid points) and PSR J1048--5832 (green dashed points).
The spectral points were obtained by performing likelihood analysis,
assuming a single PL shape in each energy bin.
The errors include both statistics and systematics.
Compared to the single PL modeling for 0.1-100 GeV or the single CPL fitting for 0.1-10 GeV (black dot-dashed lines), the CPL+PL combination model (each component and their sum; blue dotted lines) represents the 0.1-100 GeV spectrum of 1FGL J1045.2--5942 best.
(b) The same spectrum as (a) but with X-ray data points of $\eta$ Car.
While the optically-thin thermal spectra below 10 keV by the ${\it Swift}$ XRT were observed during the last periastron passage simultaneously with ${\it Fermi}$,
the ${\it INTEGRAL}$ and ${\it Suzaku}$ observations sensitive to the hard tail emission were carried out before the ${\it Fermi}$ launch
\citep{integral,sekiguchi}.}
\label{fig2}
\end{figure}

The resulting 0.1--100 GeV spectra of
1FGL J1045.2--5942 and PSR J1048--5832
are shown in Figure~\ref{fig2} (a).
The integrated photon flux ($>$ 100 MeV) and spectral parameters of PSR J1048--5832 are obtained as 
$F(> 100 {\rm MeV}) = 2.6^{+0.3}_{-0.4} \times 10^{-7}$ photons s$^{-1}$ cm$^{-2}$, ${\it \Gamma} = 1.4 \pm 0.2$ and $E_{\rm cutoff} = 2.0^{+0.6}_{-0.4}$ GeV
with additional systematic uncertainties of $\sim$10\%,
and consistent with the values obtained by \citet{psrj1048}.
Compared with the good CPL fit of PSR J1048--5832,
the spectrum of 1FGL J1045.2--5942 has a similar convex shape below 10 GeV,
but shows a hard emission tail above 10 GeV.
The 10--100 GeV emission of 1FGL J1045.2--5942 is detected at a significance level
 of $> 11\sigma$ (and 6$\sigma$ even including the systematic uncertainty),
while for PSR J1048--5832 only upper limits are obtained above 10 GeV.
This result is consistent with that in Figure~\ref{fig1},
where 1FGL J1045.2--5942 is clearly brighter than PSR J1048--5832 in the $>$ 10 GeV map.
Therefore, the 100 MeV--100 GeV spectrum cannot be fitted by either the single PL or CPL model,
and is represented well by their combination (CPL+PL model).
As the obtained physical parameters are summarized in Table~\ref{tab_fit},
the log likelihood ($L$) differences of $-2 \times \Delta$ log $(L)$, corresponding to the test statistic (TS) value,
for a single CPL and the CPL+PL model with respect to
the single PL are 4 and 68, respectively.
From the CPL+PL model, we obtain
${\it \Gamma} = 1.6 \pm 0.2$ and $E_{\rm cutoff} = 1.6^{+0.8}_{-0.5}$ GeV
for the low-energy CPL model,
and ${\it \Gamma} = 1.9^{+0.2}_{-0.3}$ for the high-energy PL component.
The fluxes below/above 10 GeV are measured as
$F$(0.1--10 GeV) $= 3.7^{+0.3}_{-0.1} \times 10^{-7}$ photons s$^{-1}$ cm$^{-2}$
and
$F$($>$ 10 GeV) $= (8 \pm 2) \times 10^{-10}$ photons s$^{-1}$ cm$^{-2}$,
and the total energy flux (0.1--100 GeV) is obtained as
$(2.4 \pm 0.1) \times 10^{-10} + (0.4 \pm 0.1) \times 10^{-10} = (2.8 \pm 0.2) \times 10^{-10}$ ergs s$^{-1}$ cm$^{-2}$,
where the systematic uncertainties are $\sim$10\% and 20\% below and above 10 GeV, respectively.
Therefore, the obtained photon flux ($>$ 100 MeV) by the ${\it Fermi}$ LAT is consistent with that reported by ${\it AGILE}$ \citep{agile}.

There is one unidentified EGRET source, 3EG J1048--5840 $(\alpha, \delta) = (162.14, -58.68)$ with the 95\% error circle of $0{\degr}.17$ \citep{3eg}, close to the positions of 1FGL J1045.2--5942 and PSR J1048--5832, as shown in Figure~\ref{fig1}.
The summed flux of 1FGL J1045.2--5942 and PSR J1048--5832 is
$\sim 6.3 \times 10^{-7}$ photons s$^{-1}$ cm$^{-2}$,
and is consistent with that of 3EG J1048--5840
[$(6.2 \pm 0.7) \times 10^{-7}$ photons s$^{-1}$ cm$^{-2}$].
Therefore, it is plausible that the EGRET source was the sum of the two LAT sources \cite[see also][]{psrj1048}.

In Figure~\ref{fig2} (b) we compare the high-energy $\gamma$-ray spectrum of
1FGL J1045.2--5942
to X-ray spectra of $\eta$ Car.
The X-ray telescope (XRT) onboard the ${\it Swift}$ satellite \citep{xrt} observed
$\eta$ Car every 2 days during the last periastron passage,
where the X-ray flux below 10 keV decreased continuously
more than one order of magnitude.
We analyzed the archival clean data in the standard way
and show two 0.5--10 keV spectra of
the first observation on 2008 December 19--20
(the highest flux; obsid 00031308002) and
the last observation on 2009 January 14
(the lowest flux; obsid 00031308009)
as the representatives.
There are prominent Fe-K${\alpha}$ lines at 6--7 keV
coming from the optically-thin thermal plasma,
and the spectral shape as well as the variability
 is similar to that observed in the periastron passage \citep{hamaguchi}.
Above 10 keV, there are three observations
by the ${\it INTEGRAL}$ and ${\it Suzaku}$ satellites
to report detection of the X-ray hard tail from $\eta$ Car
up to 50--100 keV
almost free of contamination from surrounding sources
\citep{integral,sekiguchi}.
We also plot the flux of the hard tail emission which is represented
by a single PL model
without the contribution of the soft thermal emission
 dominated below 10 keV.
Although the observations were performed before the launch of
the ${\it Fermi}$ satellite and not simultaneous
and the variation of the hard X-ray emission is still uncertain
along the orbit,
the $\gamma$-ray spectrum of 1FGL J1045.2--5942 is smoothly connected
to the X-ray spectra of $\eta$ Car with the photon index of $\sim$2,
which is consistent with theoretical prediction of
inverse-Compton emission from the CWB system \citep[e.g.,][]{cwb2}.

\subsection{Timing Analysis}

X-ray observations have detected intense variability from $\eta$ Car in 2008--2009 
 \citep{rxte_lc},
and \citet{agile} also reported a large $\gamma$-ray flare on 2008 October 11--13 on a time scale of 2 days by ${\it AGILE}$.
To study the association between the $\gamma$-ray source and $\eta$ Car,
it is very important to search for coincident variability in their light curves.
Therefore, we generated the 0.1--100 GeV light curve of 1FGL J1045.2--5942
from the 1-year LAT dataset
by aperture photometry with an aperture of 0$\degr$.5
 (using $\tt gtbin$ and $\tt gtexposure$ tools).
We perform this analysis on a time scale of 2 days
to compare with the result of \citet{agile} who used the same time scale.

Figure~\ref{fig3} shows the resulting 2-day light curve,
where the contribution of the Galactic and extragalactic diffuse emission
within this aperture of $3.2 \times 10^{-7}$ photons s$^{-1}$ cm$^{-2}$
that is obtained from the spectral analysis is subtracted.
Our light curve does not show any evidence for flaring activity
on a 2-day time scale.
The deviation from constant flux is $\chi^2/dof = 169/175 \sim 1$ 
and can be fully attributed to statistical fluctuations.
The largest 2-day flux amounts to $(9 \pm 3) \times 10^{-7}$ photons s$^{-1}$ cm$^{-2}$
which present a $\sim 2 \sigma$ deviation from the average value of
$3.7 \times 10^{-7}$ photons s$^{-1}$ cm$^{-2}$.
Although the average flux is consistent with that obtained by ${\it AGILE}$ as mentioned in $\S$~2.1,
we do not detect the large $\gamma$-ray flare of $(27.0 \pm 6.5) \times 10^{-7}$ photons s$^{-1}$ cm$^{-2}$ (6--10 times brighter than the average)
that has been reported by \citet{agile},
which if it existed,
should have been easily detected by the $Fermi$ LAT.

To compare the LAT data with the ${\it AGILE}$ results more directly,
Figure~\ref{fig3a} shows the counts maps around 1FGL J1045.2--5942 with the same 2-day binning
as Fig.~2 of \citet{agile},
where the LAT map labeled ``Oct 11--13'' covers the epoch of the large flare by ${\it AGILE}$
(i.e., from 2008 October 11th 02:27 to 13th 04:16 UT).
As shown in Figure~\ref{fig1}, the average flux at PSR J1028--5819 is the highest in this region ($\sim$30\% brighter than that at 1FGL J1045.2--5942).
This trend is also the same for all the four 2-day maps
and is consistent with the result obtained in the above light curve,
although there are statistical fluctuations
due to the limited number of detected events in the short exposure.

Because the ${\it AGILE}$ satellite provides good sensitivity to $\sim$100 MeV $\gamma$-rays,
the flare reported by \citet{agile} may only have occurred at low energies
where the LAT effective area drops rapidly with the ``diffuse'' event selection
that was used in this work \citep{lat_performance}. 
We therefore also perform a light curve analysis in the 100-400 MeV energy band,
for which we employ a 2$\degr$ extraction radius
to take into account the increase in size of the point spread function at lower energies.
Table~\ref{tab_eventnumber} summarizes the number of events detected
around the position of 1FGL J1045.2--5942 by the LAT in each period.
In addition to photons from 1FGL J1045.2--5942,
those events include photons from diffuse Galactic and extragalactic emission
and the two nearby pulsars PSR J1048--5832 and PSR J1028--5819.
We calculate that the mean number of events from 1FGL J1045.2--5942 in each time period
is $\sim$6 counts ($\sim$1/7 of the total) and $\sim$5 counts ($\sim$1/2 of the total)
for the 100--400 MeV band within 2$\degr$ and 0.1-100 GeV within 0.5$\degr$,
respectively.
These numbers are statistically consistent among the four periods,
and we conclude that the flux of 1FGL J1045.2--5942 does not vary significantly over the entire week
in not only the total 0.1--100 GeV band but also the softer 100--400 MeV band.
If the large flare reported by \citet{agile} occurred within the field of view of the LAT and the flux of 1FGL J1045.2--5942 increased by 6--10 times higher than the average in only the soft energy band around 100 MeV,
the total number of the events detected by the LAT in 100--400 MeV should have increased by $\sim$35--60 counts (a factor of 2--3), which corresponds to the 5--9 $\sigma$ increase.
Again we conclude that this should have been easily detected by the ${\it Fermi}$ LAT.

In Figure~\ref{fig3b} we plot the LAT effective area within the radius of 2$\degr$ around 1FGL J1045.2--5942 in 0.1--100 GeV and 100--400 MeV obtained by $\tt gtexposure$ around the time of the ${\it AGILE}$ large flare,
as well as the off-axis angle from the LAT boresight to the source.
The tool $\tt gtexposure$ calculates the spectrally-weighted average exposure (in the unit of ``cm$^2$ s'') also taking into account the truncation of the point spread function by the finite size of the aperture toward the source region.
To obtain the effective area (``cm$^2$'') for both 0.1--100 GeV and 100--400 MeV bands,
we performed $\tt gtexposure$ assuming the photon index of 2.0 with the time binning of 60 s and the 2$\degr$ radius,
and divided the output by 60 s.
The absolute value of the 100--400 MeV area is about half of that of the total 0.1--100 GeV band, and large enough to detect the flare as described above with the actual number of the detected events.
Since ${\it Fermi}$ was operating in survey mode,
1FGL J1045.2--5942 was observed in a repeating pattern every two orbits ($\sim$3 hours),
with a duration of $\sim$50 minutes in the first orbit
at boresight angles between 0$\degr$ and 70$\degr$ ($=$ the size of the LAT field of view),
and with a duration of $\sim$20 minutes in the second orbit
at the boresight angle only around 70$\degr$.
The LAT observed the source near the edge of the field of view in the second orbit,
and the effective area toward the source was not very efficient during the time.
Therefore, we note that the time coverage of $\eta$ Car
by both satellites is not identical,
and the large flare reported by ${\it AGILE}$ could have escaped detection by the ${\it Fermi}$ LAT
if it lasted for only a few hours.
However, restricting the AGILE detected flare duration to $\sim$2 hours would imply that the source flux increased by approximately 120--220 times the average.

To search for $\gamma$-ray variability coincident with the X-ray variability of $\eta$ Car,
which occurs on time scales of a few months,
we also obtained the LAT light curves with 1-week binning
for 1FGL J1045.2--5942,
and for comparison also PSR J1048--5832 and PSR J1028--5819.
We analyzed the 0.1--10 GeV events during every 7 days
by the same likelihood method described in $\S$ 2.1,
but fixing the photon index and cutoff energy (i.e., spectral shape)
of each source at its average values
and keeping only the flux as a free parameter.
For all 1FGL J1045.2--5942, PSR J1048--5832 and PSR J1028--5819, the CPL model was used.
This method, fitting all the detected events within the ROI of $10\degr$,
models the contamination from other sources and the diffuse background,
and measures the fluxes of
all the three sources simultaneously.
In each week, 1FGL J1045.2--5942 is detected with a TS value of at least 10 (typically $\sim$40), and
the resulting light curves are shown in Figure~\ref{fig4} as well as the 2--10 keV X-ray light curve of $\eta$ Car \citep{rxte_lc} obtained by the Proportional Counter Array (PCA) onboard \textit{Rossi X-ray Timing Explorer} \citep{jsg+96}.
The deviations of the $\gamma$-ray light curves from the constant fluxes are $\chi^2/dof = 43/50$ (the probability to be consistent with the mean is $P$ = 75\%) for 1FGL J1045.2--5942, $41/50$ ($P$ = 81\%) for PSR J1048--5832 and $53/50$ for PSR J1028--5819 ($P$ = 36\%),
where we applied 3\% systematic uncertainty derived in the ${\rm Fermi}$ LAT first source catalog \citep{lat_1yr}.
Since PSR J1048--5832 and PSR J1028--5819 are thought to be stable sources
and there is no significant deviation in their light curves.
We conclude that
1FGL J1045.2--5942 does not show any significant variability on the time scale of $>$ 1 week, either.
Above 10 GeV, the number of events detected from 1FGL J1045.2--5942 integrated over the current 1-year observation amounts to only $\sim$30, which makes a variability analysis in this band unfeasible.

Subtracting the contamination from surrounding sources
in the $RXTE$ PCA field of view,
the optically-thin thermal X-ray emission below 10 keV from $\eta$ Car itself varies
by $\sim$ two orders of magnitude
around the periastron passage \cite[Figure~\ref{fig2} and ][]{hamaguchi}.
In contrast, the obtained $\gamma$-ray light curve of 1FGL J1045.2--5942
is rather steady even during the last periastron passage of $\eta$ Car
in 2009 January (Figure~\ref{fig4}),
with the possible fluctuation of
at most $\pm 2 \times 10^{-7}$ photons s$^{-1}$ cm$^{-2}$
or a factor of $\sim$3.



To investigate the possible variation of 1FGL J1045.2--5942
coincident with the X-ray light curve of $\eta$ Car,
we next accumulated the LAT events into three periods to increase the statistics.
Referring to the X-ray light curve of $\eta$ Car, the $\gamma$-ray data are divided as shown in Figure~\ref{fig4};
the X-ray flux of $\eta$ Car increased in Period 1,
decreased by more than an order of magnitude in Period 2
and returned to the average in Period 3.
From the 0.1--10 GeV likelihood analysis of each period in the same way as that in $\S$~2.1 (i.e., all the spectral parameters are fitted freely),
the $\gamma$-ray source 1FGL J1045.2--5942 was detected significantly
in each period with TS values of 750, 231 and 830, respectively.
Their spectra and the obtained parameters are shown in Figure~\ref{fig5} and Table~\ref{tab1}, respectively.
We also performed the same 0.1--10 GeV analysis for the average data and determined the spectral shape as shown in Figure~\ref{fig2} (a), to compare the results one another including the average spectrum.
The power-law index and cutoff energy in all the periods are consistent with the values determined for the total dataset.
This result confirms that 1FGL J1045.2--5942 is detected
close to the average flux
even during the X-ray faintest state of $\eta$ Car (Period 2),
and the $\gamma$-ray variability (if it exists) is very small (at most a factor of $\sim$1.5)
compared with that in X-ray ($>$ one order of magnitude).

The similarity of the spectra of 1FGL J1045.2--5942 and PSR J1048--5832 below 10 GeV may suggest that 1FGL J1045.2--5942 could be associated
with a yet unknown pulsar.
However, using our 1-year dataset, we searched for possible pulsations by the blind search technique \citep{blindsearch} like other LAT pulsars, and 
did not detect any pulsations on time scales from several 
tens of milliseconds to 1 day.
Considering the number of events detected from 1FGL J1045.2--5942 is $\sim$5000,
the upper limit of the pulsed flux is estimated as $\sim$10\% level of the 1FGL J1045.2--5942 flux, i.e., as $\sim 3 \times 10^{-8}$ photons s$^{-1}$ cm$^{-2}$.

Given the pulsar-like spectrum of 1FGL J1045.2--5942, we also searched
for X-ray pulsations in an archival ${\it RXTE}$
observation of the Anomalous X-ray Pulsar 1E~1048.1$-$5937 with          
1FGL J1045.2--5942 in the 1$\degr$ field of view              
\citep{jmr+06}.  We chose an observation        
from 1996 September 29--30, spanning 99\,ks with 53\,ks of exposure      
(Obsid 10192-03), and selected events from the top Xenon detection        
layer, from all 5 Proportional Counter Units (PCUs), in the              
energy range $\sim$2--10\,keV.
After each photon was barycentered using        
the position of 1FGL J1045.2--5942,
time series were created
with 1/512\,s time resolution.
We then performed a                
Fast Fourier Transform and searched for a peak                          
in the power spectrum not harmonically related to the bright            
Anomalous X-ray Pulsar (P=6.452\,s).
As a result, we found no evidence for a pulsed signal in the X-ray band, either.
According to the                          
time resolution of the time series and strong red noise in the          
power spectrum, the obtained result is
sensitive to pulsations with frequencies between
10\,ms and 100\,s.
Following the methods in \citet{van89} and \citet{vvw+94},
the 99\% confidence                                                
upper limit on a sinusoidal pulsed signal                               
is estimated to be 0.14 count s$^{-1}$ PCU$^{-1}$
corresponding to $3.1 \times 10^{-13}$\,ergs\,s$^{-1}$\,cm$^{-2}$ (2-10\,keV),
where a simple power-law model                    
with a spectral index of 2.0 and an absorption column density of
$1 \times 10^{21}$\,cm$^{-2}$ are assumed.

\section{Discussion}

Our analysis of the 1st year ${\it Fermi}$ LAT dataset revealed a $\gamma$-ray source (1FGL J1045.2--5942) that is spatially consistent with the location of $\eta$ Car.
The average flux ($>$ 100 MeV) is measured as $3.7^{+0.3}_{-0.1} \times 10^{-7}$ photons s$^{-1}$ cm$^{-2}$, and consistent with that reported by ${\it AGILE}$ \citep{agile}.
The $\gamma$-ray signal is detected significantly throughout the LAT energy band up to $\sim$100 GeV.
The 0.1--100 GeV energy spectrum cannot be fit by either a single PL or CPL model,
but is represented well by a combination of a cutoff power-law model ($<$ 10 GeV) and a hard power-law component ($>$ 10 GeV).
As one of the most powerful colliding wind binary systems that is known, $\eta$ Car is certainly a promising candidate for the underlying source of 1FGL J1045.2--5942.
Although the hard X-ray observations were not performed simultaneously with ${\it Fermi}$,
the measured $\gamma$-ray spectrum is smoothly connected to the hard X-ray emission
above 10 keV from $\eta$ Car with the photon index of $\sim$2,
which is consistent with the theoretical study of CWBs,
namely the spectrum from the inverse-Compton emission
 by the interaction between
stellar photons and accelerated particles
at the colliding-wind regions \citep[e.g.,][]{cwb2},
yet so far we have no convincing evidence (such as correlated time variability or flaring events) that allows a physical link to be established.
In particular, our 2-day light curve does not show any indication for short duration $\gamma$-ray flares
in either softer 100--400 MeV or total 0.1--100 GeV bands
 and we are unable to confirm the large $\gamma$-ray flare on 2008 October 11--13 reported by \citet{agile}.

As blazars are the dominant source population of the GeV sky, we cannot exclude the possibility that 1FGL J1045.2--5942 is associated with a background blazar in the direction of the Carina nebula. Three arguments, however, disfavor such an hypothesis. Firstly, the probability of finding at least one background blazar within the 95\% confidence region of 1FGL J1045.2--5942 amounts to only $7 \times 10^{-5}$ \citep{lat_1yr}. We estimate this probability from the source density of $\sim$90 high-latitude GeV sources per sr (which mostly are blazars) with 1--100 GeV fluxes of $\gtrsim 1 \times 10^{-9}$ photons s$^{-1}$ cm$^{-2}$ and from the solid angle of about $8 \times 10^{-7}$ sr of the 95\% confidence error region. Secondly, 1FGL J1045.2--5942 appears to be a stable source in contrast to many blazars that are time variable \citep{lat_blazar}. Thirdly, the spectrum of 1FGL J1045.2--5942 is rather complex, combining an exponentially cut-off power-law spectrum with an additional hard power-law tail at high energies. None of the blazars observed by ${\it Fermi}$ shows similar spectral characteristics.

On the other hand, the spectrum is qualitatively very similar to that observed from the Crab \citep{crab}. While for the Crab, the exponentially cutoff power-law is attributed to the pulsar, the high-energy power-law component is attributed to the pulsar wind nebula. Since the LAT has recently discovered many $\gamma$-ray bright but radio and X-ray faint or quiet pulsars \citep{pulsar}, a pulsar and its nebula may represent a reasonable explanation for the observed emission, although no pulsar or pulsar wind nebula is yet known near $\eta$ Car. However, $\eta$ Car is a member of the young and massive open star cluster Tr 16 that may well house some young and energetic pulsars, similar to PSR J2032+4127 which has been recently discovered by the ${\it Fermi}$ LAT toward the Cyg OB2 massive star cluster \citep{pulsar,radio_pulsar}, and may be associated with the unidentified TeV source TeV J2032+4130 \citep{tev_unid} that may represent the corresponding pulsar wind nebula.
Alternatively, the exponentially cutoff power-law and the hard power-law components could arise from two physically unassociated sources, although the positions obtained in the $<$ 10 GeV and $>$ 10 GeV bands are close enough to be consistent within the current statistical uncertainty.

In any case, the study of the variability of 1FGL J1045.2--5942 will be key to identify the underlying source. Were 1FGL J1045.2--5942 to be related to $\eta$ Car, some modulation of the $\gamma$-ray flux is expected during the 5.54 yr long orbit \citep{cwb2} and the ${\it Fermi}$ LAT is ideally suited to reveal this modulation.
On the other hand, were 1FGL J1045.2--5942 a pulsar,
the pulsation might be detected below the current upper limits in $\gamma$-ray and X-ray,
and help may also come from a deep radio
search that is able to unveil young and energetic pulsars. Furthermore, deep observations in the TeV domain by ${\it HESS}$ and ${\it CANGAROO}$,
and in the future ${\it CTA}$ may reveal the related pulsar wind nebula if it exists.


\if
From the analysis of the 1-year $Fermi$ LAT dataset
for a $\gamma$-ray source 1FGL J1045.2--5942 discovered close to $\eta$ Car,
the position of 1FGL J1045.2--5942 is evaluated 
 at $(RA, Dec) = (161.265, -59.695)$ with the 95\% statistic error of $0{\degr}.031$,
resulting in that $\eta$ Car locates well within the error circle
and other bright X-ray sources are excluded
from the candidates of the association.
The light curve with 2-day binning does not confirm
the presence of the large $\gamma$-ray flare in 2008 October
reported by $AGILE$ \citep{agile}.
As the counterpart of 1FGL J1045.2--5942,
the following three candidates are considered;
(1) really associated with $\eta$ Car,
(2) a background blazar
and (3) a pulsar in the Carina nebula.

The spectrum of 1FGL J1045.2--5942 has a convex-concave shape
below/above 10 GeV
and is represented by the combination of
the soft CPL model and the hard PL component
with the total 0.1--100 GeV energy flux of $(2.8 \pm 0.2) \times 10^{-10}$ ergs s$^{-1}$ cm$^{-2}$
with the additional systematic uncertainty of $\sim$10\%.
Although the flux matches those of typical blazars observed by the LAT \citep{lat_blazar},
the spectral shape is different from usual simple PL or CPL modeling of blazars.
Therefore, the association with a background blazar is not likely for 1FGL J1045.2--5942.
On the other hand, this spectrum with the soft CPL and hard PL components
is very similar to those of pulsars with pulsar wind nebulae.
Since the LAT has recently discovered $\gamma$-ray bright but radio/X-ray faint pulsars \citep[e.g.,]{cta1}
,
pulsars may become a candidate,
although there are previously no known pulsars and diffuse nebula emission (except for the bipolar nebula of $\eta$ Car) around the position of 1FGL J1045.2--5942 determined by the LAT.
However, if the variation of Period 2 and 3 investigated in $\S$~2.2
is really the celestial behavior,
1FGL J1045.2--5942 could not be such temporary stable sources.

In the case when 1FGL J1045.2--5942 is actually associated with $\eta$ Car,
it would be the first situation to detect the high-energy $\gamma$-rays from the CWB.
However,
it means that the $\gamma$-ray emission from the CWB
is not always proportion to the X-ray
during the periastron passage,
since the 0.1--10 GeV emission is clearly detected
with almost the average brightness
even at the X-ray faintest state of $\eta$ Car (Period 3)
when the X-ray luminosity decreases by $\sim$ two orders of magnitude \citep{hamaguchi}.
On time scales of the 7-day and longer binning (Period 1--4),
the light curve of 1FGL J1045.2--5942 is almost steady,
and the $\gamma$-ray variability is at most by a factor of $\sim$2
during this year.
Therefore, to clear whether 1FGL J1045.2--5942 is originated from
$\eta$ Car or a pulsar,
it is important to continuously search for significant variation
(coinciding with other wavelengths)
through the continued LAT all-sky survey.
\fi

\acknowledgments

\begin{figure}
\includegraphics[width=0.7\textwidth]{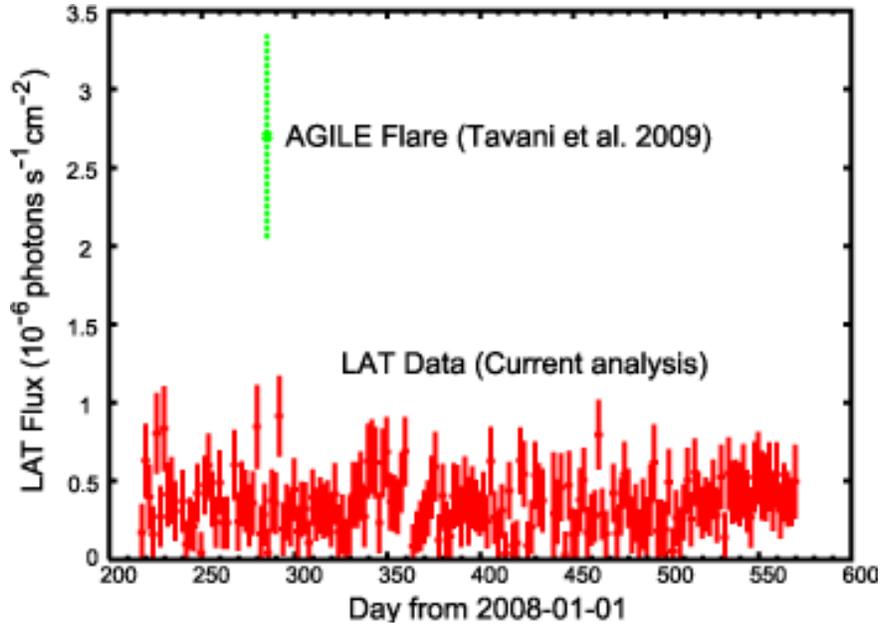}
\caption{Light curve of 1FGL J1045.2--5942 by aperture photometry
with the binning of 2 days.
Contribution of Galactic and extragalactic diffuse emission is already subtracted.
Periastron passage of $\eta$ Car is around the day 360 measured from 2008 January 1st.
We also show the $\gamma$-ray flux reported by \citet{agile} at day 285.
}
\label{fig3}
\end{figure}

\begin{figure}
\includegraphics[width=0.7\textwidth]{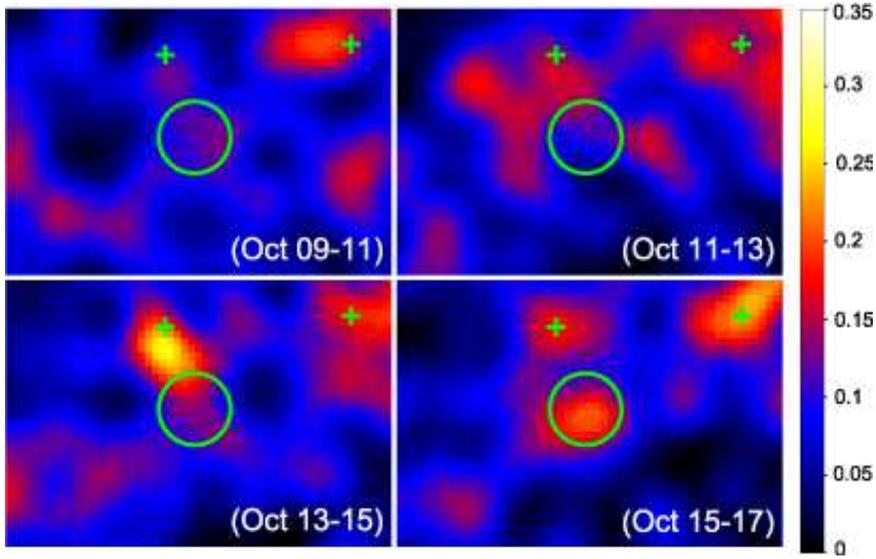}
\caption{The same as the 0.1--10 GeV counts map shown in Figure~\ref{fig1} but for the time around the large flare reported by ${\it AGILE}$ during 2008 October 11--13, corresponding to Fig.~2 of \citet{agile}.
Each counts map includes all of the events in the 0.1--100 GeV band with 6-bin Gaussian smoothing.
The time interval in the panel labeled ``Oct 11-13'' is exactly
the same as that of the ${\it AGILE}$ flare
from 2008 October 11th 02:27 to 13th 04:16 (UT),
while the time intervals in the remaining panels amount to 2 days.
The variation of the features among the different panels
is primarily explained by statistical fluctuations
due to the limited photon statistics in these short time intervals.
The green circle represents the 0${\degr}$.5 radius region around 1FGL J1045.2--5942,
within which the LAT light curve in Figure~\ref{fig3} is derived.
The two crosses indicate the positions of the two pulsars PSR J1048--5832 and PSR J1028--5819.
}
\label{fig3a}
\end{figure}

\begin{figure}
\includegraphics[width=0.7\textwidth]{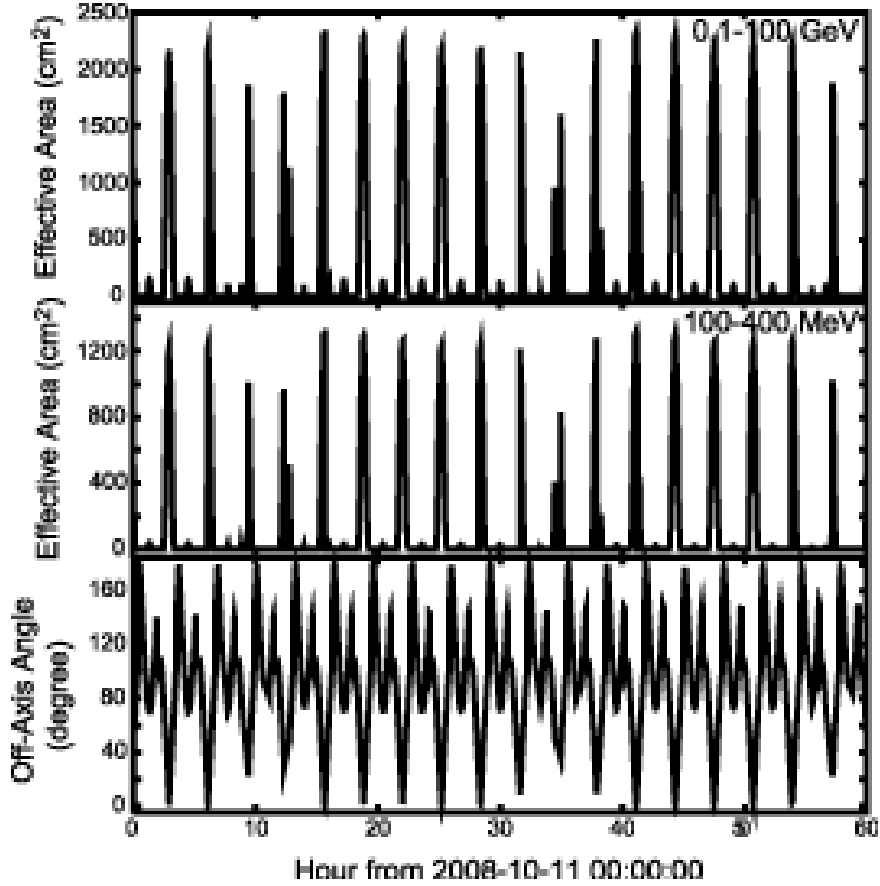}
\caption{
Effective area of 1FGL J1045.2--5942 in 0.1--100 GeV and 100--400 MeV energy bands by the LAT during the ${\it AGILE}$ large flare.
Off-axis angle from the LAT boresight to the source is also plotted at the bottom.
The time binning of each plot is 60 s.
According to the survey operational mode, the LAT periodically observed 1FGL J1045.2--5942 twice with the duration of $\sim$50 and 20 minutes every 3 hours.
See the text for more details.
}
\label{fig3b}
\end{figure}

\begin{figure}
\includegraphics[width=0.7\textwidth]{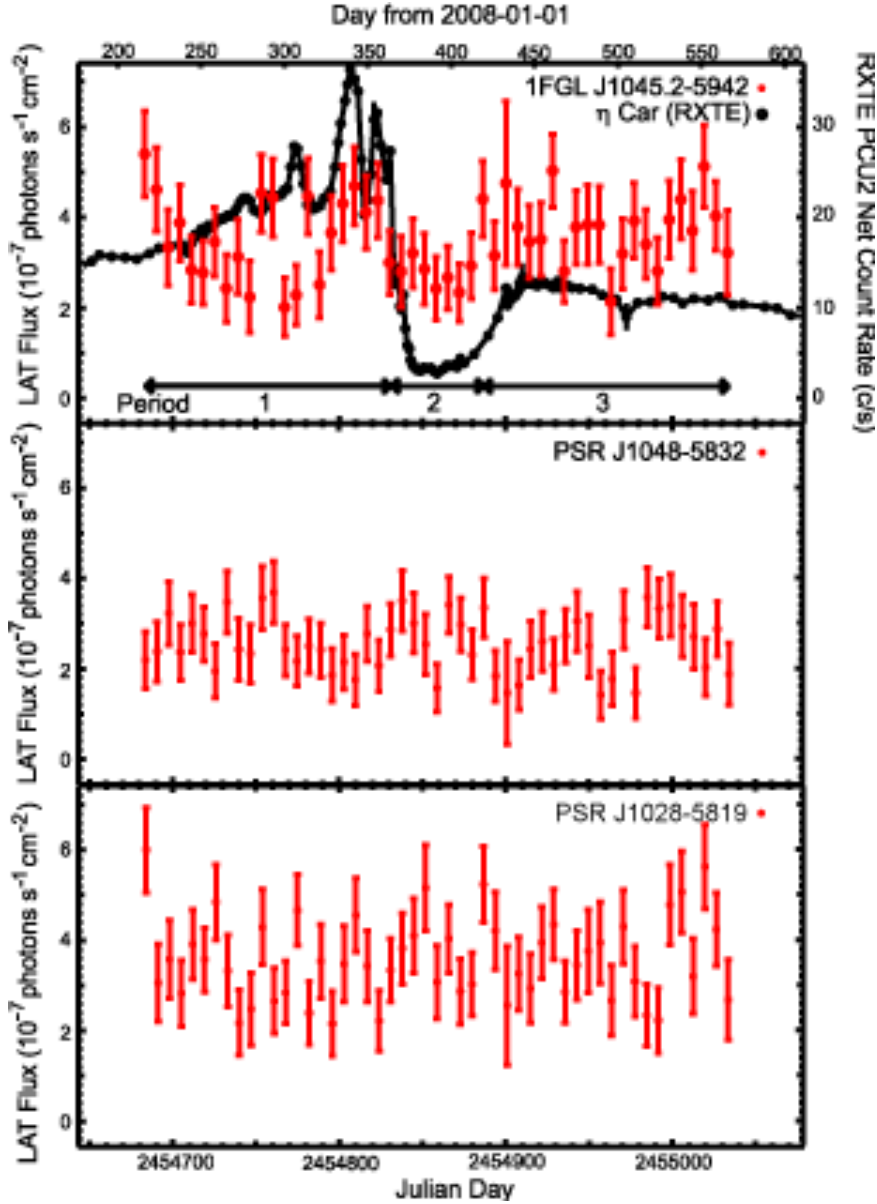}
\caption{
Light curves of 1FGL J1045.2--5942, PSR J1048--5832 and PSR J1028--5819 obtained by
likelihood analysis.
Each bin is one-week long, and the flux is in the 0.1--10 GeV energy band.
The 2--10 keV X-ray light curve of $\eta$ Car measured by the ${\it RXTE}$ PCA is shown overlaid in the top panel \citep{rxte_lc}.}
\label{fig4}
\end{figure}

\begin{figure}
\includegraphics[width=0.7\textwidth]{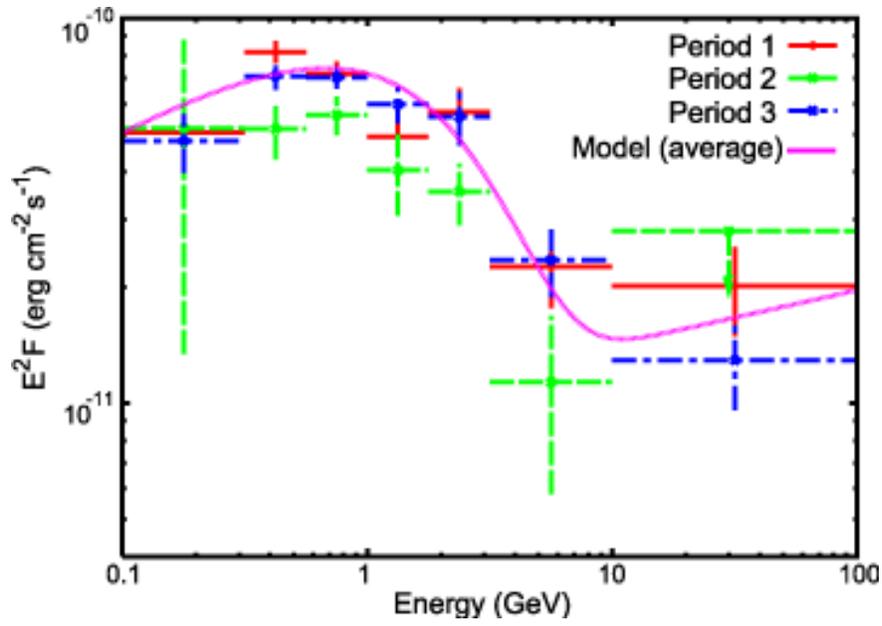}
\caption{The same as Figure~\ref{fig2} (a) but for that of each Period in Figure~\ref{fig4}.
The best-fit model for the average spectrum is also shown (solid line).}
\label{fig5}
\end{figure}

The ${\it Fermi}$ LAT Collaboration acknowledges support from a number of agencies and institutes for both development and the operation of the LAT as well as scientific data analysis.
These include NASA and DOE in the United States, CEA/Irfu and IN2P3/CNRS in France, ASI and INFN in Italy, MEXT, KEK, and JAXA in Japan, and the K.\ A.\ Wallenberg Foundation, the Swedish Research Council and the National Space Board in Sweden.
Additional support from INAF in Italy and CNES in France for science analysis during the operations phase is also gratefully acknowledged.
We thank M. Corcoran for providing the X-ray light curve of $\eta$ Car by the ${\it RXTE}$ PCA in Figure~\ref{fig3}.



{\it Facilities:} \facility{Fermi (LAT)}.




\clearpage



\clearpage


\clearpage


\clearpage


\clearpage


\clearpage

\clearpage

\clearpage

\begin{table}
\begin{center}
\caption{Spectral fitting results of the average spectrum of 1FGL J1045.2--5942 in the 0.1-100 GeV energy band.$^{\rm a}$}
\begin{tabular}{lcccc}
\hline
Model & Photon index & Cutoff & Photon flux ($>$ 100 MeV) & TS value$^{\rm b}$\\
 & & (GeV) & ($10^{-7}$ photons s$^{-1}$ cm$^{-2}$) & \\
\hline
PL & $2.2 \pm 0.1$ & $\cdots$ & $4.6^{+0.5}_{-0.3}$ & $\cdots$ \\
CPL & $2.1 \pm 0.1$ & $> 13 $ & $4.4^{+0.2}_{-0.1}$ & 4 \\
CPL+PL & $1.6 \pm 0.2$ & $1.6^{+0.8}_{-0.5}$ & $3.2^{+0.3}_{-0.5}$ & \\
 & $1.9^{+0.2}_{-0.3}$ & $\cdots$ & $0.5^{+0.9}_{-0.3}$ & 68 \\
\hline
\multicolumn{5}{@{}l@{}}{\hbox to 0pt{\parbox{160mm}{\footnotesize
$^{\rm a}$ All errors are statistical only. Systematic errors should be additionally taken into account with 10\%, 5\% and 20\% at 100 MeV, 560 MeV and 10 GeV, respectively.\\
$^{\rm b}$ The TS value relative to the PL model fitting.
}\hss}}
\label{tab_fit}
\end{tabular}
\end{center}
\end{table}

\begin{table}
\begin{center}
\caption{The number of $\gamma$-ray events detected around the position of 1FGL J1045.2--5942 by the LAT in two selection cases (100--400 MeV in a 2$\degr$ radius, and 0.1--100 GeV in 0$\degr$.5) around the date of the ${\it AGILE}$ flare (October 11--13).$^{\rm a}$}
\begin{tabular}{lcc}
\hline
Date & 100--400 MeV \& 2$\degr$$^{\rm b}$ & 0.1--100 GeV \& 0$\degr$.5 \\
\hline
October 09--11  & 34 & 9 \\
October 11--13  & 44 & 8  \\
October 13--15  & 42 & 10 \\
October 15--17  & 46 & 13 \\
\hline
\multicolumn{3}{@{}l@{}}{\hbox to 0pt{\parbox{160mm}{\footnotesize
$^{\rm a}$ Events include the contribution from both source and diffuse background \\ (Galactic and extraglactic) emission.\\
$^{\rm b}$ There are also contributions from PSR J1048--5832 and PSR J1028--5819 \\ within the radius of 2$\degr$.
}\hss}}
\label{tab_eventnumber}
\end{tabular}
\end{center}
\end{table}


\begin{table}
\begin{center}
\caption{Physical parameters of the CPL model obtained for the average and each Period spectra of 1FGL J1045.2--5942 in the 0.1--10 GeV band.$^{\rm a}$}
\begin{tabular}{lcccc}
\hline
Period & Photon index & Cutoff & Photon flux ($>$ 100 MeV) & Energy flux (0.1--10 GeV)\\
 & & (GeV) & ($10^{-7}$ photons s$^{-1}$ cm$^{-2}$) & ($10^{-10}$ ergs s$^{-1}$ cm$^{-2}$) \\
\hline
Average & $1.8 \pm 0.1$ & $3.3^{+1.5}_{-0.7}$ & $3.7^{+0.3}_{-0.1}$ & $2.4 \pm 0.1$ \\
1 & $1.8^{+0.1}_{-0.2}$ & $3.3^{+2.2}_{-1.1}$ & $3.9^{+0.2}_{-0.3}$ & $2.5 \pm 0.1$ \\
2 & $1.9^{+0.2}_{-0.3}$ & $3.3^{+6}_{-1.6}$ & $3.4^{+0.3}_{-0.4}$ & $2.0 \pm 0.1$ \\
3 & $1.8 \pm 0.2$ & $3.5^{+3.0}_{-1.1}$ & $3.9^{+0.4}_{-0.2}$ & $2.6 \pm 0.1$ \\
\hline
\multicolumn{5}{@{}l@{}}{\hbox to 0pt{\parbox{160mm}{\footnotesize
$^{\rm a}$ All errors are statistical only. Systematic errors should be additionally taken into account with 10\%, 5\% and 20\% at 100 MeV, 560 MeV and 10 GeV, respectively.\\
}\hss}}
\label{tab1}
\end{tabular}
\end{center}
\end{table}

\end{document}